# Large-scale wireless network management via Open-RAN Tandem Apps: Cell on/off switching use case

Paweł Kryszkiewicz[1,2], Łukasz Kułacz[1,2], Marcin Pakuła[1,2], Marcin Dryjanski[2], Marcin Hoffmann[1,2], Piotr Skrzypczak[1,2], Heiko Lehmann[3], Martin Stahn[3]

[1]Poznan University of Technology, Poland; [2]Rimedo Labs, Poland; [3]T-Labs, Germany

**Abstract**

With growing mobile-network complexity, management and optimization have become increasingly difficult. Centralized algorithms face high control-data overhead and computational load, while distributed approaches often perform far from optimally. The O-RAN architecture introduces two tiers of RAN Intelligent Controllers (RICs), enabling hierarchical network-management schemes. This work proposes Tandem Apps: a pair of tightly coupled optimization mechanisms running on both controllers. We show how to design Tandem Apps through architectural and functional splitting to achieve an agile, low-complexity solution that still preserves a global network view. As an example, we implement Tandem Apps for cell on/off switching and evaluate them in a large heterogeneous network using real network data. Although the Tandem Apps concept is new, it remains fully compliant with the O-RAN standard, as validated using commercial network software.

## I. Introduction

With the advent of 6G technology [1], the wireless network will become even more complex and heterogeneous with thousands of Base Stations (BSs) from different vendors, utilizing various radio access technologies (RANs) ranging from 2G to 6G, carrier frequencies, and transmit power. This creates a significant challenge in an online network reconfiguration based on the current traffic needs.

One of the solutions for this problem is the O-RAN [2] that, thanks to the openness of communication interfaces, allows for unified configuration of multiple Base Stations (BSs), composed of O-RAN Radio Units (O-RUs), O-RAN Distributed Units (O-DUs), and O-RAN Centralized Units (O-CUs). The latter two are called E2 Nodes. The entities that allow for continuous network monitoring and reconfiguration are Near Real Time RAN Intelligent Controller (Near-RT RIC) and Non Real Time RIC (Non-RT RIC) that operate on different time scales, and run goal-oriented applications [3], called xApps and rApps, respectively.

The O-RAN architecture facilitates solving many non-trivial network optimization problems. One of these is the network's adaptation to the changing traffic, temporally or spatially, to increase energy efficiency (EE). There are multiple energy-saving (ES) features provided by O-RAN [4], e.g., RF channel switching or small timescale base station sleeping, called Advanced Sleep Mode. However, the most investigated approach, even before the O-RAN era [5], is Cell ON/OFF switching (COOS). It provides high energy savings as nearly all cell components can be turned off. However, most existing COOS solutions are tested in simplified network models, e.g., the bitrate is calculated using Shannon's capacity formula, and full channel state information is assumed. Moreover, many algorithms optimize a set of active BSs at a given time instance, not considering continuous network operation, e.g., [6]. There are also some recent solutions utilizing O-RAN architecture for COOS. In [7], the authors propose xApps that turn off radio channels if the load is below an arbitrarily chosen threshold. However, the algorithm cannot turn on inactive radio channels if the load increases. Moreover, the solution is tested in a relatively small, homogenous network composed of 12 O-RUs. An even more recent example is a COOS xApp utilizing Deep Reinforcement Learning [8]. Unfortunately, probably because of exponentially rising with network size actions space, this is also tested in a homogeneous, 7 BSs network without spatially or temporally varying traffic. Therefore, the scalability of the proposed solution is limited.

This paper demonstrates that the two-tier Non-RT and Near-RT RIC architecture, running the proposed Tandem Apps, i.e., tightly coupled rApps and xApps pursuing a common goal, enables scalable, continuous-time network management even in large, heterogeneous networks.

Sec. II outlines the gap between optimal centralized optimization and the need for real-time, continuous control, introducing the Tandem Apps concept within the O-RAN architecture and providing design guidelines for effective rApp–xApp coordination. Sec. III presents the Cell ON/OFF Switching Tandem App, which significantly reduces energy consumption while preserving user QoS. A key strength of our work is its evaluation using real BS deployments and traffic from a live, multi-frequency Deutsche Telekom network, detailed in Sec. IV. Sec. V validates the feasibility of deploying Tandem Apps in an operational O-RAN environment through a Proof-of-Concept using Keysight RICtest and the Juniper RIC, with conclusions and future work in Sec. VI.

## II. O-RAN architecture for hierarchical network management

### A. Need for scalable network control

With the increasing complexity and size of wireless networks, efficient resource management has become challenging, where the solutions' scalability gains paramount importance. Scalability involves three key aspects in control algorithm design: network size/complexity, temporal dynamics, and computational or signaling load. Network size may reflect added complexity, e.g., such as activating more antennas or carriers, or spatial expansion through deploying new cells and serving more users. The time dimension affects the system in several ways. First, the network itself changes over time, e.g., through fluctuations in the number of UEs, so the control actions must be able to adapt to varying conditions. Moreover, the algorithm should limit its memory by forgetting old data, unapplicable to the current network configuration. Finally, network reconfigurations conducted over short time windows, i.e., a "snapshot optimization," may lead to unstable network configurations. A scalable solution should limit these variations. From the perspective of computational complexity and signaling load, a good algorithm, as defined, e.g., for Cell Free Massive MIMO [9], scales linearly with the number of network nodes, implying fixed complexity per cell.

### B. Network control architecture for scalable network optimization

To meet these requirements, an appropriate solution architecture is needed. Three classical approaches to decision-making in network optimization are centralized, distributed, and hybrid. In the centralized approach, all network state information is collected in one place and decisions are made centrally. While this can yield optimal decisions, it incurs high complexity and signaling overhead and scales poorly. In contrast, in a distributed approach, nodes make decisions independently based on limited, typically local information. This improves scalability and reduces complexity, but may lead to suboptimal solutions.

Considering the above aspects, a scalable network optimization solution should avoid fully centralized decision-making, yet also not rely on a fully distributed approach to achieve high performance. This suggests adopting a hybrid or hierarchical solution that combines the advantages of both paradigms to address scalability.

### C. O-RAN Tandem Apps

A hierarchical algorithmic architecture naturally fits the O-RAN framework, where short- and long-timescale decisions, and local versus global control, are split between the Near-RT RIC and Non-RT RIC. These controllers host xApps and rApps, respectively, which perform network optimization.

We propose Tandem Apps, a coordinated pair of an rApp and xApp with complementary, non-overlapping functions working toward a shared optimization goal, such as reducing energy consumption via COOS. The overall architecture is shown in Fig. 1. Tandem Apps follow O-RAN specifications: the rApp on the Non-RT RIC sets policies, gathers analytics, and can train AI models using large datasets. Operating on timescales of tens of seconds to minutes, it captures long-term trends[10] and provides global, potentially complex optimization over large network areas. It supplies master parameters, enrichment information, or AI models to the xApp and governs its behavior through policies.

The Near-RT RIC, integrated with the RAN and connected to E2 nodes, hosts xApps that collect gathers the Key Performance Measurements (KPM) and execute fast control actions. Operating between 10 ms and 1 s, the xApp enables near-instantaneous reconfiguration, such as triggering serving-cell changes during handover[10]. Its short timescale limits computational load and scope, allowing multiple local xApps to operate under rApp coordination.

In large networks with strict latency and signaling constraints, Tandem Apps offer agility through fast, lightweight xApps while preserving a global, long-term view via the rApp, all using limited control information.

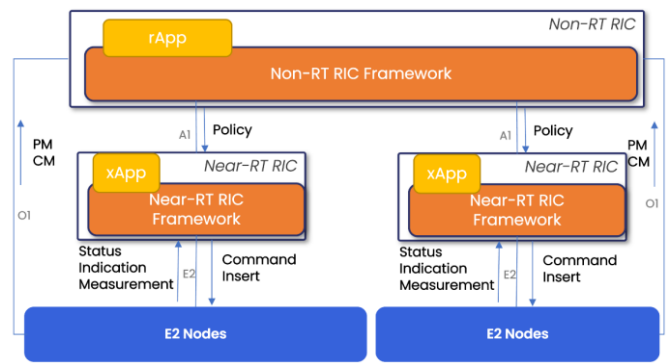

*Figure 1 Tandem Apps in O-RAN Architecture*

### D. Apps coupling & Tandem Apps design recommendations

Tandem Apps represent a specific form of App interaction within O-RAN. In practice, many Apps may run simultaneously with different objectives and control variables. Their interactions can be

grouped as follows:

- Uncorrelated (or disjoint) Apps: They modify network parameters independently and do not affect each other, e.g., two handover Apps operating hundreds of kilometers apart, where the uncorrelation will be enforced by the distance. Ideally, App designers should strive for minimal coupling to ease multi-vendor integration.
- Negatively coupled Apps: Their goals conflict, causing parameters oscillations or contradictory actions, e.g., a COOS xApp switching cells off while a throughput-maximization xApp switches them on. Conflict-mitigation mechanisms can address this [11].
- Positively coupled Apps: The system benefits from synergy between Apps pursuing a common goal, even if indirectly, through different metrics. Tandem Apps intentionally create such positive coupling. Another example is COOS combined with Traffic Steering: COOS reduces the number of active BSs, lowering the denominator of EE metric, while traffic steering boosts user rates, jointly improving EE.

This classification is continuous rather than binary, i.e., multiple levels of positive/negative coupling are possible. Designing effective Tandem Apps is therefore non-trivial. However, we propose several general recommendations, supported by the Tandem Apps example in the next section, to guide scalable future designs, namely:

- Both Apps in a Tandem pair should pursue the same overall goal, even if they use different metrics, to avoid contradictory actions and network oscillations.
- There should be a clear hierarchy: one App handles network reconfiguration, while the other provides the policies or high-level decisions. This avoids conflicts when both operate asynchronously modifying the network.
- Because xApps and rApps operate on different timescales, timers are needed to block new actions until previous changes are fully applied and network metrics have stabilized. These timers should differ for the fast, local xApp and the global, slower rApp.
- The Tandem Apps algorithm should be split between the xApp and rApp with minimal control-message exchange. The xApp should handle low-complexity tasks to enable fast, agile reactions to network conditions.

Due to current O-RAN limitations, such as the lack of mechanisms for Apps to understand each other's goals, Tandem Apps will likely need to come from a single vendor.

## III. Cell ON/OFF switching in O-RAN by Tandem Apps

After outlining the general Tandem Apps concept, we detail it using a COOS use case in a large, heterogeneous network, an area of high importance for Mobile Network Operators (MNOs). The key Apps: Tandem Apps composed of COOS xApp and COOS rApp, and a separate Traffic Steering (TS) xApp, and accompanying information exchange are shown in Figure 2. While the wireless network is a complex system, the COOS algorithm cannot be discussed in a vacuum, independently of other network functions. Crucial here is the TS functionality included in Figure 2 as a separate App, not belonging to the COOS tandem. For example, when COOS-xApp asks for switching off a cell, TS must move the affected UEs to other cells to maintain Quality of Service (QoS). Both the delay of this operation and the choice of target cells can significantly affect COOS performance and may cause QoS degradation [4]. The TS may be handled directly by the RAN (within the E2 Node) or by a dedicated TS xApp. While COOS could, in principle, include TS internally, we assume here that the COOS tandem only learns TS behavior from network statistics. Thus, the proposed COOS tandem is TS-agnostic and adapts to whichever TS solution is deployed.

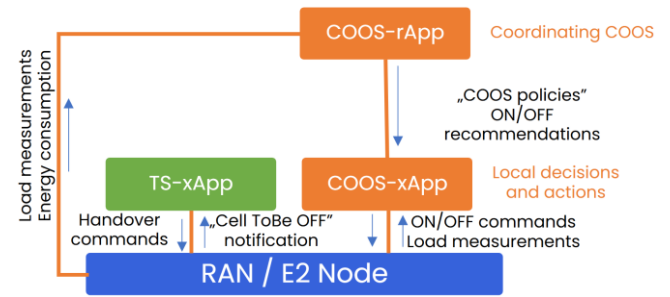

*Figure 2 Cell On/Off Switching Tandem Apps architecture; TS: Traffic Steering*

The COOS Tandem Apps consist of two components: the COOS-xApp and COOS-rApp. The xApp collects load information from its controlled cells and, using global policies from the rApp, decides on turning on/off cells. A cell is turned off when its load falls below a given threshold ($\alpha_{off}$) and turned on when the average load of its neighbors exceeds a given threshold ($\alpha_{on}$). After any state change, the cell and its neighbors are blocked for a fixed period to prevent ping-pong effects. The xApp issues decisions to the E2 Node as „cellToBeOff" message. Because suitable thresholds are difficult to derive analytically, the rApp provides them. The lower $\alpha_{off}$, the smaller the probability of turning off a cell. The higher $\alpha_{on}$, the lower the probability of turning on a cell. The thresholds should balance energy savings and UEs

QoS. For this purpose, the rApp can be used, as it has long-term and wide-area network awareness.

The COOS-rApp coordinates all actions and controls the xApp through the A1 interface, connecting Near-RT RIC with Non-RT RIC. Based on the system KPIs observation, the rApp adjusts thresholds ($\alpha_{off}$, $\alpha_{on}$) used by COOS-xApp to accomplish long-term objectives of the network, e.g., keeping users' outage within a target range. Its broader spatial and temporal view allows for increasing or decreasing insensitivity of cells switching on or off depending on daytime, but also emergency situations. A simplified rApp algorithm monitors the average system outage ($\beta_{sys}$), and a ping-pong (PP) flag equal to one if some cells performed two state changes over a short period, zero otherwise. Its goal is to maximize the number of switched-off cells while keeping the required outage (defined as a range by the MNO) and a low number of cell state changes (to control the ping-pong effect).

The algorithm can be divided into 5 cases:
1. If the average system outage $\beta_{sys}$ exceeds the outage target range, indicating too few active cells, and PP is one, signaling a ping-pong effect, we first decrease the $\alpha_{off}$ to reduce the intensity of cell switch-offs;
2. Otherwise, if the average system outage $\beta_{sys}$ is greater than the target outage range, indicating too few active cells, and some cells are still off, $\alpha_{on}$ is decreased to increase cell switching on intensity;
3. Otherwise, if the average system outage $\beta_{sys}$ is below the target outage range, meaning more cells could be turned off, and the flag PP is one, indicating excessive PP, the cell activation threshold $\alpha_{on}$ is increased to lower cell switching on intensity;
4. Otherwise, if the average system outage $\beta_{sys}$ is lower than the allowed value range, and there are cells that could be switched off, increment $\alpha_{off}$ to increase cell switching off intensity;
5. otherwise, keep $\alpha_{on}$ and $\alpha_{off}$ unchanged.

The TS-xApp helps COOS-Tandem Apps in the operation. The TS-xApp subscribes to cell state changes. Before the cell turn-off happens, the RAN notifies the TS-xApp to take care of cleaning up a given cell (i.e., moves users out of this cell by invoking handover commands). Once empty, RAN switches off the cell and notifies both the COOS-xApp and TS-xApp of the taken action.

Intentionally, the above approach does not use any sophisticated AI tools, to keep full explainability of its operation, important for MNOs. Though in the future, other tools can be considered.

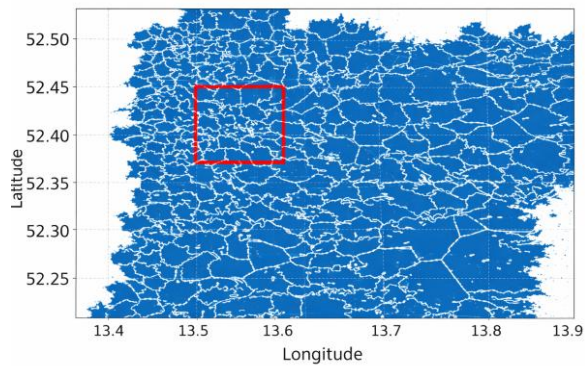

Figure 3 Coverage of 5G cells running at 773 MHz carrier.

## IV. Large-scale network algorithm testing

The proposed Tandem Apps was evaluated in a realistic, large-scale, heterogeneous network scenario thanks to the network topology and users' traffic patterns obtained from a real Deutsche Telekom network. The coverage areas for cells at 773 MHz carrier frequency are shown in Figure 3. While the spatial scalability of the Tandem Apps was confirmed in this network, for long-term simulations, requiring high computational complexity caused by emulating each single UE, a limited network area was selected (red rectangle), still larger and more heterogeneous than considered in previous works [7][8]. The area includes 60 cells distributed across 13 base stations operating at three different carrier frequencies: 15 cells at 773 MHz (coverage layer, cells that cannot be turned off), 39 cells at 2160 MHz (capacity layer), and 6 cells at 3655 MHz (capacity layer). The number of users and their required throughput were characterized statistically, separately in 100 m x 100 m areas (called "pixels") every 30 minutes. Each pixel was obtained based on a 24-hours dataset, allowing for modeling of significant spatial and temporal variability in the network. The processing of data was similar to the one presented in [12]. In the simulation, in each pixel, new users are generated according to the Poisson point process, with a 2 minutes mean service time, and an exponentially distributed throughput. Therefore, the algorithm is not "learning" UEs as each run results in a random UEs realization. However, the parameters of random processes were selected in such a way that the mean throughput and number of users have aligned with measurements both temporally and spatially. The simulated network consists of 5346 pixels and covers a 6.5 x 8.0 km area. The average number of active UEs per pixel varies from 0.01 to 0.21. Therefore, in some pixels during 24h period, only a few UEs are created, while there exists a pixel of 164 users appearing, on average. Similarly, the mean required throughput per pixel, varies between pixels from 0.56 Mbps to 19 Mbps. This shows that the traffic requirements vary significantly both spatially and temporally. Per user throughput variability over time and space is also visible. More insight into the utilized dataset can be obtained by checking [13].

The simulator uses the location of cells and their parameters, like transmit power, height, tilt, frequency, bandwidth, etc., based on a real network deployment. The UEs' locations are modeled using a random waypoint model with a pedestrian speed. The BS-UE path loss is modeled using Urban Macro or Urban Micro propagation models, depending on the BS location, extended with spatially correlated shadowing [14]. To facilitate a large number of UE simulations, a truncated Shannon formula has been used for bitrate estimation, as suggested for system-level modeling of 5G systems. The BS power consumption uses the 3GPP model [15]. Most importantly, in this paper, the *outage*, used in rApp and reported as KPI below, is defined as a percentage of network UEs that do not achieve the full throughput requested by a given UE. A UE in outage typically still is connected to the network. However, the specific KPI used for this purpose is up to the MNO.

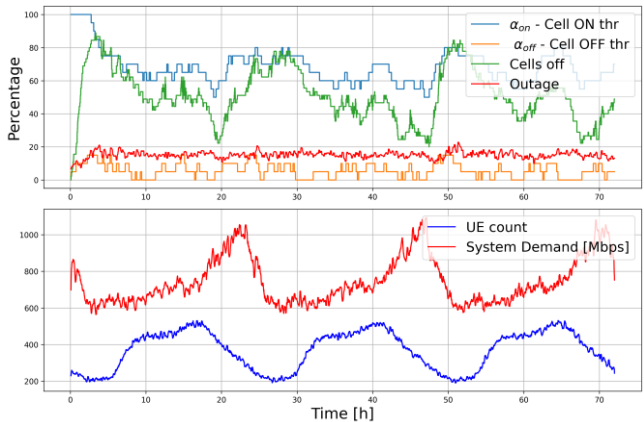

*Figure 4 Network statistics and rApp dynamic thresholds over a few days' simulation.*

First, in Fig. 4, network statistics over a few days of network operation are presented, including the COOS thresholds provided to the xApp from the rApp. The network starts with all cells active and initial COOS thresholds stopping xApp from any cell's state change, i.e., $\alpha_{off}=0\%$ and $\alpha_{on}=100\%$. As average system outage $\beta_{sys}$ is below the target outage range, i.e., 15%±1%, the rApp increases $\alpha_{off}$ to 5% (cells with load lower than 5% can be turned off). Recall that, according to the above definition of outage, this is not equivalent to 15% of UEs not served by the network. Turning off cells starts at this moment and continues until around 80% of the capacity-layer cells are turned off. This results in network outage exceeding the goal range, making the rApp decrease $\alpha_{on}$ (starting around the 5[th] hour), allowing for turning ON some cells. Observe, the xApp itself decides which cells should be activated based on the load conditions of the neighbors of inactive cells, being the lower tier of decision-making. Over the next hours, the $\alpha_{off}$ and $\alpha_{on}$ thresholds are dynamically adjusted in response to the network state, keeping the number of inactive capacity cells between 20 and 80% while being close to the target outage range. Most importantly, even though the traffic demand changes in time and space, as visible in the number of users and total required throughput on the bottom plot, the Tandem Apps maintains the stability of the network. Still, future improvement of algorithm parameters, e.g., $\alpha_{on}$ or $\alpha_{off}$ step size, is possible to allow for adjustment to network dynamics without decisions overshooting. Moreover, a very advantageous feature of the proposed Tandem Apps is scalability. We have observed that the xApp decisions, i.e., turning ON/OFF cells, that can be decentralized and performed locally at xApps responsible for particular network parts, are around 5 times more often than the rApp decisions. While it is anticipated that rApp should be centralized, its reduced workload significantly helps the solution's scalability.

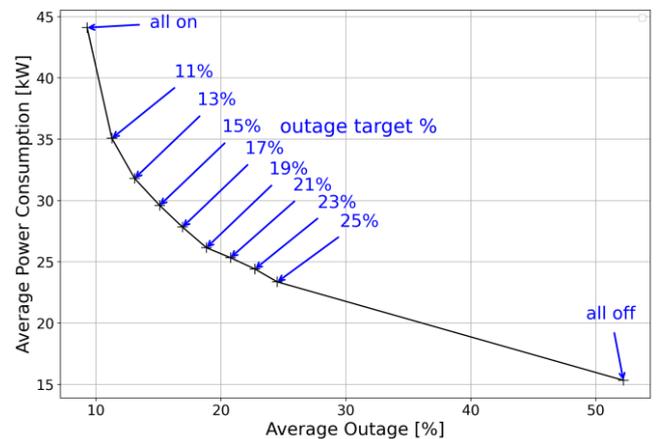

*Figure 5 Network power consumption vs outage while varying the rApp goal outage (in blue).*

The proposed COOS Tandem Apps provides to the MNO a high-level control. In Figure 5 the average power consumption and outage from the whole 3 days of the network simulation are shown for different COOS rApp outage goals (in blue). Reference results for all cells active (no energy saving) and all capacity-layer cells turned off are also marked. It shows that the proposed COOS Tandem Apps can find a balance between the users' QoS, and the energy savings. Most interestingly, there is initially a rapid decrease in energy consumption if the network outage is increased slightly, i.e., allowing the outage to increase by 4 percentage points, from 9.3% to 13.1% on x axis, results in around 28% energy consumption decrease, from 44.09 kW to 31.79 kW on y axis.

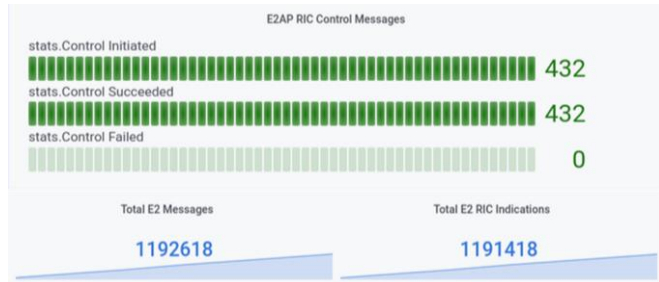

*Figure 6 E2 interface message exchange statistics from RICTest by Keysight Technologies*

## V. Proof of Concept (PoC)

Finally, the COOS xApp/rApp tandem has been integrated with commercial RIC platforms provided by Juniper Networks and evaluated using the O-RAN network emulator provided by Keysight Technologies. While the former section was focused on the performance evaluation of the algorithm itself, the PoC demonstrates the potential for commercial deployment of the proposed Tandem Apps by validating it against the O-RAN protocol stack, exploiting E2, and O1 interfaces emulated in real time by the Keysight RICTest software, R1, and A1 interfaces being an internal part of Juniper Networks' RIC platform. In other words, the existing O-RAN standard allows for its implementation. However, regarding the A1 interface implementation of Tandem Apps requires defining a custom policy going beyond the data types strictly defined by the O-RAN ALLIANCE. This could suggest that some future discussion on more flexible approach to A1 interface is needed. The scenario for the PoC, compared to the other publicly showcased in O-RAN ALLIANCE, assumed the use of a large-scale network with 15 BSs, and a total number of 39 cells, deployed based on the real-world data from Deutsche Telekom. While the implementation confirmed the ability to perform the COOS process effectively, similar as shown in the previous section for simulations, below we focus on the control messages exchanged over O-RAN interfaces to assess the solution's signaling scalability.

While the O-RAN specifications define Non-RT control loop as above 1 s, our implementation experience is that the O1 Performance Metrics required by the rApp generates relatively slow traffic of tens of reports per 1 minute. This constitutes even stronger justification of placing long term optimization in Non-RT RIC. The same applies to the A1 interface, where once per 5 minutes, a single policy containing thresholds ($\alpha_{off}$, $\alpha_{on}$) is sent from COOS-rApp in Non-RT RIC to COOS-xApp in Near-RT RIC. More control messages go through the E2 interface connecting COOS-xApp and RAN. The E2 statistics obtained over 1 hour of COOS xApp/rApp operation are summarized in Fig. 6.

During one hour of operation, a total of 1,192,618 E2 messages were sent, resulting in about 331 E2 messages per second. A small portion of this amount is setup messages sent by BSs to initiate E2 connection, and subscription requests sent by both TS and COOS xApps to configure reception of necessary input data. More of the exchanged E2 messages (a total of 432), where the control actions. These are both requests to turn on/off a cell, initiated by the COOS-xApp, and handover commands initiated by the TS-xApp. The biggest signaling overhead is related to E2 indication messages (a total of 1,191,418), containing mostly cell load values required by the COOS-xApp, and received signal strength as an input to the TS-xApp. However, the latter ones constitute the larger portion of E2 indication messages, as in this case, they are reported by each user once per second.

These statistics prove the scalability of the proposed Tandem Apps. The Non-RT RIC has low signaling overhead on both O1 and A1 interfaces, making it possible for the COOS-rApp to serve thousands of cells. On the other hand, while the COOS xApp obtains a significant number of RAN reports, its computational complexity is low. Moreover, multiple instances of Near-RT RIC can be deployed to cover sub-areas and provide a fast reaction to the changes in network load.

## VI. Conclusions

This paper introduced the Tandem Apps concept for managing large, heterogeneous O-RAN networks through a hierarchical RIC architecture. Simulations based on real-network data show that the approach provides significant performance gains in the cell on/off switching scenario, and the Proof of Concept using O-RAN-compliant commercial tools confirms that it can be deployed with existing interfaces, reducing time-to-market.

Future work will include evaluating the COOS Tandem App in networks with different traffic characteristics, such as IoT-dominated environments, considering additional algorithm fine-tuning. Another direction is the use of Machine Learning methods, including Reinforcement Learning, to adapt thresholds using richer contextual information. Finally, extending the study beyond the COOS use case will help better understand the strengths and limitations of the Tandem

Apps concept and the extent to which current O-RAN standards support it.

## Acknowledgment

The work was funded by T-Labs (Deutsche Telekom)

## Literature

[1] C. X. Wang, et al., "On the Road to 6G: Visions, Requirements, Key Technologies, and Testbeds," IEEE Communications Surveys & Tutorials, vol. 25, no. 2, pp. 905-974, Q2 2023

[2] M. Polese, L. Bonati, S. D'Oro, S. Basagni, T. Melodia, "Understanding O-RAN: Architecture, Interfaces, Algorithms, Security, and Research Challenges," IEEE Communications Surveys & Tutorials, 2023, Vol. 25, No. 2, pp. 1376-1411

[3] M. Hoffmann, et al., "Open RAN xApps Design and Evaluation: Lessons Learnt and Identified Challenges," IEEE J. on Selected Areas in Comm., vol. 42, no. 2, pp. 473-486, Feb. 2024,

[4] L. Kundu, X. Lin, R. Gadiyar, "Toward Energy Efficient RAN: From Industry Standards to Trending Practice," IEEE Wireless Communications, vol. 32, no. 1, pp. 36-43, Feb. 2025

[5] J. Wu, et al., "Energy-Efficient Base-Stations Sleep-Mode Techniques in Green Cellular Networks: A Survey," IEEE Communications Surveys & Tutorials, Q2 2015,

[6] E. Oh, K. Son and B. Krishnamachari, "Dynamic Base Station Switching-On/Off Strategies for Green Cellular Networks," IEEE Trans. on Wireless Comm., vol. 12, no. 5, May 2013

[7] X. Liang, et al., "Enhancing Energy Efficiency in O-RAN Through Intelligent xApps Deployment," WINCOM, Leeds, UK, 2024.

[8] Bordin, Matteo, et al. "Design and Evaluation of Deep Reinforcement Learning for Energy Saving in Open RAN." IEEE CCNC 2025.

[9] Demir ÜT, BjËrnson E, Sanguinetti L (2021), "Foundations of User-Centric Cell-Free Massive MIMO". Foundations and Trends in Signal Processing, Vol. 14 No. 3-4 pp. 162–472

[10] A. Kliks, et al., *"Towards Autonomous Open Radio Access Networks"*, ITU Journal on Future and Evolving Technologies, June 2023

[11] C. Adamczyk, A. Kliks, "Conflict mitigation framework and conflict detection in O-RAN Near-RT RIC." IEEE Comm. Magazine 2023

[12] D. Rose, J. Baumgarten, T. Kurner, "Spatial Traffic Distributions for Cellular Networks with Time Varying Usage Intensities Per Land-Use Class," IEEE VTC, Vancouver, Canada, 2014.

[13] Rimedo Labs & T-Labs. *"Multi-scale hierarchical rApp-xApp tandem for energy saving using real mobile network data."* MWC 2025. https://tinyurl.com/zj9w4c7t

[14] C. Zhang, et al., "Two-dimensional shadow fading modeling on system level," IEEE PIMRC, Sydney, Australia, 2012

[15] M. Oikonomakou, et al., "A Power Consumption Model and Energy Saving Techniques for 5G-Advanced Base Stations," IEEE ICC Workshops, Rome, Italy, 2023.

**Pawel Kryszkiewicz** is an Associate Professor at the Institute of Radiocommunications, PUT, and the Technical Director of Rimedo Labs. He is the author of over 100 publications on multicarrier systems, green communications, and Open RAN.

**Łukasz Kułacz** is an R&D expert at Rimedo Labs, and an Assistant Professor at the Institute of Radiocommunications, PUT. His research field includes mainly dynamic spectrum management, spectrum sharing, and O-RAN applications.

**Marcin Pakuła** is an R&D Engineer at Rimedo Labs. His research interests include software-defined radio and O-RAN architecture development.

**Marcin Dryjanski (SM)** serves as CEO and principal consultant at Rimedo Labs. He received his Ph.D. from the Poznan University of Technology in 2019. He is a co-author of a book and many articles on Open RAN, 5G, and beyond.

**Marcin Hoffmann** is a Technical Solution Manager at Rimedo Labs and a Ph.D. candidate at PUT. His research interests include machine learning for 5G/6G network management.

**Piotr Skrzypczak** is a Junior R&D Engineer at Rimedo Labs with research focusing on conflict mitigation in Open RAN.

**Heiko Lehmann** is Tribe Lead for Cybersecurity and Digital Twin at T-Labs, Deutsche Telekom. He was founding Thematic Area Leader for "Smart Energy" at the European Institute of Technology and Innovation. He has published extensively in theoretical physics, informatics, engineering, and business topics.

**Martin Stahn** is a research architect at T-Labs, Deutsche Telekom. He is working on mathematical modelling and optimization in the context of multi-level network simulation. He has published in mathematics, namely probabilistic numeric and operator-based analysis and optimization of complex dynamical system